\documentclass[conference]{IEEEtran}

\usepackage[T1]{fontenc}
\usepackage{amsmath}
\usepackage{booktabs}
\usepackage{tabularx}
\usepackage{graphicx}
\usepackage{stfloats}
\usepackage{microtype}
\usepackage{needspace}
\usepackage{siunitx}
\usepackage{tikz}
\usetikzlibrary{arrows.meta,calc,decorations.pathmorphing,fit,patterns,positioning}

\definecolor{lfiink}{HTML}{1D2028}
\definecolor{lfielectron}{HTML}{216FC4}
\definecolor{lfihole}{HTML}{D95C2B}
\definecolor{lfin}{HTML}{B8D4EE}
\definecolor{lfip}{HTML}{F5E3DA}
\definecolor{lfibeam}{HTML}{8850C7}
\definecolor{lfigate}{HTML}{747B89}

\tikzset{
  lfiwire/.style={draw=lfiink,line width=1.15pt,line cap=round,line join=round},
  lfiarrow/.style={-{Latex[length=2.5mm,width=1.7mm]},line width=1.25pt},
  lfifield/.style={draw=lfiink!65,-{Latex[length=2.1mm,width=1.4mm]},line width=.85pt},
  lfiphoton/.style={draw=lfibeam,line width=1.5pt,-{Latex[length=2.6mm]},
    decorate,decoration={snake,amplitude=2.1pt,segment length=7pt,post length=7pt}},
}
\newcommand{\lfie}[1]{\fill[lfielectron] #1 circle (.105);}
\newcommand{\lfih}[1]{\draw[lfihole,fill=white,line width=1.4pt] #1 circle (.105);}
\newcommand{\lfiplus}[1]{\draw[lfiink!50,line width=.85pt]
  ($#1+(-.055,0)$)--($#1+(.055,0)$)
  ($#1+(0,-.055)$)--($#1+(0,.055)$);}
\newcommand{\lfiminus}[1]{\draw[lfiink!50,line width=.85pt]
  ($#1+(-.055,0)$)--($#1+(.055,0)$);}
\newcommand{\lfiflash}[1]{\begin{scope}[shift={#1}]
  \foreach \a in {0,45,...,315}{\draw[lfibeam,line width=.95pt]
    (\a:.16)--(\a:.26);}
\end{scope}}

\usepackage[hidelinks]{hyperref}
\usepackage{xurl}

\IEEEoverridecommandlockouts

\newcommand{\reg}[1]{\texttt{#1}}
\newcommand{\bitname}[1]{\texttt{#1}}

\title{Shoot on (Core)Sight: Re-enabling Secure Debug on the RP2350 with
Photon-Emission Microscopy and Laser Fault Injection}

\author{
\IEEEauthorblockN{Antoine Plin}
\IEEEauthorblockA{
\textit{Ledger, Paris, France (intern)} \\
\textit{ESILV}, France \\
\textit{Georgia Institute of Technology}, USA \\
aplin3@gatech.edu}
}

\begin{document}
\raggedbottom
\maketitle

\begin{abstract}
Fault-resistant policy storage does not guarantee fault-resistant enforcement.
We demonstrated this distinction on the RP2350A4 using differential
photon-emission microscopy (PEM) and laser fault injection (LFI). PEM narrowed the area for laser scanning,
which found two sites where pulses set the \reg{DEBUGEN} bits enabling core~1 Secure debug. An online attack sequence using
only Serial Wire Debug (SWD) feedback restored sustained access. After rescue reset,
the sequence recovered a 128-bit test secret. This reset
reloads the secret's runtime page lock from a stored policy that permits
Secure reads and prevents user firmware from tightening the lock.
\end{abstract}

\begin{IEEEkeywords}
hardware security,
laser fault injection,
photon-emission microscopy,
secure debug
\end{IEEEkeywords}

\section{Introduction}
\label{sec:introduction}

Microcontrollers protect firmware and data through secure boot, hardware
isolation, and persistent access
policies~\cite{nist-ir-8259a,nist-sp-800-193,rp2350-security}. Protecting
the stored policy alone does not ensure that the hardware enforcing it is
resistant to faults. On the RP2350, the mutable \reg{DEBUGEN} register can
override \reg{CRIT1.DEBUG\_DISABLE}, a redundantly encoded one-time
programmable (OTP) flag~\cite[Secs.~3.5.9 and~13.4]{rp2350-datasheet}, so a fault in that register
can bypass the policy. Restoring
Secure debug permits an attacker to halt the processor and read firmware and
data accessible to Secure software. Laser fault injection (LFI) can alter
individual register bits, but locating the circuitry for a register whose
location on the die is unknown requires a substantial search.

Differential photon-emission microscopy (PEM) can narrow this search by
isolating light emission associated with a selected operation. Previous
studies localized CPU logic, register-file bits, and embedded
memories~\cite{nedospasov2012functional,lima2024data,perrin2025betrayed}.
We applied this approach to a weakly emitting, memory-mapped security register.
By comparing images taken while the chip was toggling different bits inside the register,
we found regions whose emission depends on those values and used them to target laser scanning.

Prior work on the RP2350 restored RISC-V debug by
faulting OTP readout at power-on~\cite{muench2025rp2350}. We instead faulted
the runtime debug controls on the RP2350A4. During offline characterization,
custom firmware executed the PEM excitation loops
and reported register values during LFI mapping. The subsequent attack used
only Serial Wire Debug (SWD) feedback.

\Needspace{6\baselineskip}
This work makes three contributions:
\begin{itemize}
  \item A differential PEM method for isolating emission associated with a
        memory-mapped register, using an atomic-XOR excitation loop,
        acquisition-order balancing, and outlier rejection.
  \item Localization of two adjacent LFI-responsive sites on one RP2350A4
        and an SWD-guided attack sequence that restores sustained core~1 Secure
        debug access.
  \item A rescue-reset bypass of runtime OTP locks, affecting any RP2350 secret those locks protect.
      Rescue mode halts boot before the firmware applies these locks, so restoring debug access
      lets us read a demo OTP secret.
\end{itemize}

\section{Background and Related Work}
\label{sec:background}

\subsection{Photon-Emission Microscopy}

Photon-emission microscopy (PEM) passively images light from operating
integrated circuits. In switching MOS transistors, carriers accelerated by
the high electric field near the drain undergo radiative transitions, producing a
broad spectrum extending into the near infrared
(NIR)~\cite{bude1992hotcarrier,phang2005review}.
In CMOS logic, this emission occurs mainly during switching, when current flows through a transistor in saturation.
Consequently, a register bit that holds a constant value contributes little light compared to a bit that toggles repeatedly~\cite{bude1992hotcarrier,phang2005review}.

Backside NIR observation images through the substrate, avoiding
front-side metal~\cite{glowacki2014backside}.

Time-integrated PEM uses an indium gallium arsenide (InGaAs) camera
and repeated execution to compensate for low photon yield. Each image, which
we call an acquisition, integrates emission over many seconds of execution.
A register bit occupies few transistors, while the processor, buses and clock tree switch many more.
Each image therefore contains emissions that are not related to the targeted logic, as well as optical and thermal background and detector noise~\cite{perrin2025betrayed}.

Differential PEM subtracts acquisitions taken while the chip runs matched
execution loops, to isolate activity associated with the operation of
interest~\cite{nedospasov2012functional,perrin2025betrayed}. Two things can
survive subtraction and must be controlled. The first is any other difference
between the loops, such as instruction pipeline or the data carried on internal
buses. Every internal value that differs between the two loops creates different switching activity, which could be mistaken for the target signal.
The second is setup drift between the two acquisitions, such as a shift in focus, alignment,
or sensor offset.

Nedospasov \emph{et al.} used PEM to localize CPU logic and status-register
flags~\cite{nedospasov2012functional}; Schl\"osser \emph{et al.} used emission
as an AES side channel~\cite{schlosser2012aes}; Lima \emph{et al.} mapped
embedded Flash and extracted stored data~\cite{lima2024data}. Perrin \emph{et al.} are the closest to our work.
They localized bits of core registers on three Cortex-M processors by building loops around the \texttt{EORS} instruction, and
demonstrated with LFI that the emission sites matched the laser-sensitive positions.
Our excitation follows the same XOR principle, but the target lies outside the CPU in a memory-mapped register of the OTP peripheral.
We also carry the localization through to an attack on a deployed debug policy,
whereas Perrin \emph{et al.} stop at matching bit positions.

PEM can narrow the spatial search, but emission during a target operation
does not establish whether a laser pulse at that location will produce the
desired fault. Local laser scans and register readback are needed to determine
which positions allow the intended bits to be modified.

\subsection{Laser Fault Injection}

\begin{figure}[t]
\centering
\newcommand{\pnslab}[2]{\begin{scope}[shift={(#1,0)}]
  \fill[lfip] (.12,1.75) rectangle (2.10,3.45);
  \fill[lfin] (2.10,1.75) rectangle (4.08,3.45);
  \path[pattern=north east lines,pattern color=lfiink!30]
    ({2.10-.33*#2},1.75) rectangle ({2.10+.33*#2},3.45);
  \draw[lfiink!55,dashed,line width=.65pt]
    ({2.10-.33*#2},1.75)--({2.10-.33*#2},3.45)
    ({2.10+.33*#2},1.75)--({2.10+.33*#2},3.45);
  \draw[lfiink!60,line width=.8pt] (2.10,1.75)--(2.10,3.45);
  \draw[lfiink!55,line width=.65pt] (.12,1.75) rectangle (4.08,3.45);
  \foreach \col in {0,...,5}{\foreach \row in {2.22,2.76,3.30}{
    \lfiminus{({.285+.33*\col},\row)}\lfiplus{({2.265+.33*\col},\row)}}}
  \pgfmathtruncatemacro{\pnlast}{5-#2}
  \foreach \col in {0,...,\pnlast}{\foreach \row in {1.95,2.49,3.03}{
    \lfih{({.285+.33*\col+.025*sin(97*\col+311*\row)},
      {\row+.02*sin(53*\col+197*\row)})}
    \lfie{({3.915-.33*\col+.025*sin(71*\col+229*\row)},
      {\row+.02*sin(89*\col+157*\row)})}}}
  \fill[lfigate] (0,2.35) rectangle (.12,2.85) (4.08,2.35) rectangle (4.20,2.85);
  \node[anchor=north west,font=\small,text=lfiink,inner sep=0] at (.12,1.63) {P};
  \node[anchor=north east,font=\small,text=lfiink,inner sep=0] at (4.08,1.63) {N};
\end{scope}}
\newcommand{\pnsource}[2]{\begin{scope}[shift={(#1,0)}]
  \draw[lfiwire] (.06,2.85)--(.06,4.20)--(2.01,4.20)
    (2.19,4.20)--(4.14,4.20)--(4.14,2.85);
  \draw[lfiwire] ({2.10+.09*#2},3.99)--({2.10+.09*#2},4.41);
  \draw[lfiink,line width=2.6pt] ({2.10-.09*#2},4.095)--({2.10-.09*#2},4.305);
  \draw[lfiink,line width=1.1pt]
    ({2.00+.30*#2},4.62)--({2.20+.30*#2},4.62)
    ({2.10+.30*#2},4.52)--({2.10+.30*#2},4.72)
    ({2.00-.30*#2},4.62)--({2.20-.30*#2},4.62);
\end{scope}}
\resizebox{0.65\columnwidth}{!}{%
\begin{tikzpicture}
\useasboundingbox (0,.62) rectangle (6.85,4.82);
\pnslab{0}{3}
\pnsource{0}{1}
\draw[lfifield,line width=1.35pt] (3.02,3.03)--(1.18,3.03);
\draw[lfiphoton] (2.10,.72)--(2.10,1.66);
\begin{scope}[shift={(2.10,1.95)},rotate=22.5,scale=.8]
  \lfiflash{(0,0)}
\end{scope}
\lfih{(1.75,1.95)}
\draw[lfihole,lfiarrow] (1.62,1.95)--(1.18,1.95);
\lfie{(2.45,1.95)}
\draw[lfielectron,lfiarrow] (2.58,1.95)--(3.02,1.95);
\lfih{(4.85,3.65)}
\node[anchor=west,font=\small,text=lfiink] at (5.15,3.65) {Hole};
\lfie{(4.85,3.15)}
\node[anchor=west,font=\small,text=lfiink] at (5.15,3.15) {Electron};
\lfiminus{(4.725,2.65)}\lfiplus{(4.975,2.65)}
\node[anchor=west,font=\small,text=lfiink] at (5.15,2.65) {Dopant ion};
\draw[lfifield] (4.60,2.15)--(5.10,2.15);
\node[anchor=west,font=\small,text=lfiink] at (5.15,2.15) {Field};
\path[pattern=north east lines,pattern color=lfiink!30,draw=lfiink!55,dashed]
  (4.64,1.48) rectangle (5.06,1.82);
\node[anchor=west,font=\small,text=lfiink] at (5.15,1.65) {Depletion};
\end{tikzpicture}
}
\caption{Reverse-biased PN junction. The junction field separates an
electron--hole pair generated by a photon in or near the depletion region,
driving the electron to N and the hole to P.}
\label{fig:pn-junction}
\end{figure}
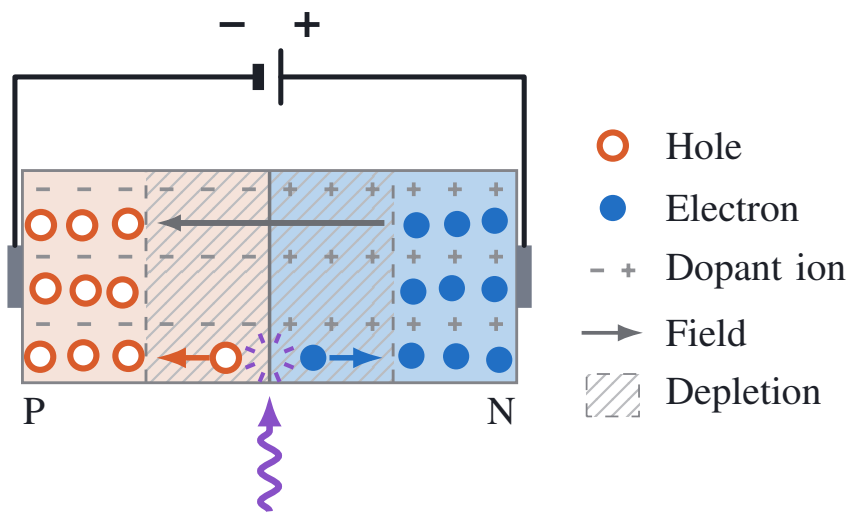

Laser fault injection (LFI) uses photons above the silicon bandgap energy to
create electron-hole pairs. Wavelengths just below the bandgap edge near \SI{1.1}{\micro\meter}
are absorbed weakly enough to penetrate the substrate yet still
generate carriers in the active layer.
Charge collection near reverse-biased junctions causes transient
photocurrents that perturb logic or stored
state~\cite{skorobogatov2002optical,roscian2013fault}
(Figure~\ref{fig:pn-junction}).
The fault depends on position, pulse timing, duration, energy, spot size, and
circuit state. Finding suitable parameters can produce repeatable, bit-selective
faults~\cite{courbon2014adjusting,champeix2015seu}.

A coarse spatial scan may miss small sensitive regions, while high pulse
energy can disturb multiple nodes rather than produce a selective bit change~\cite{courbon2014adjusting} or cause a latch-up that can leave the
target unresponsive and prevent further injections~\cite{selmke2018locked}.

\subsection{RP2350 Security and Debug Architecture}

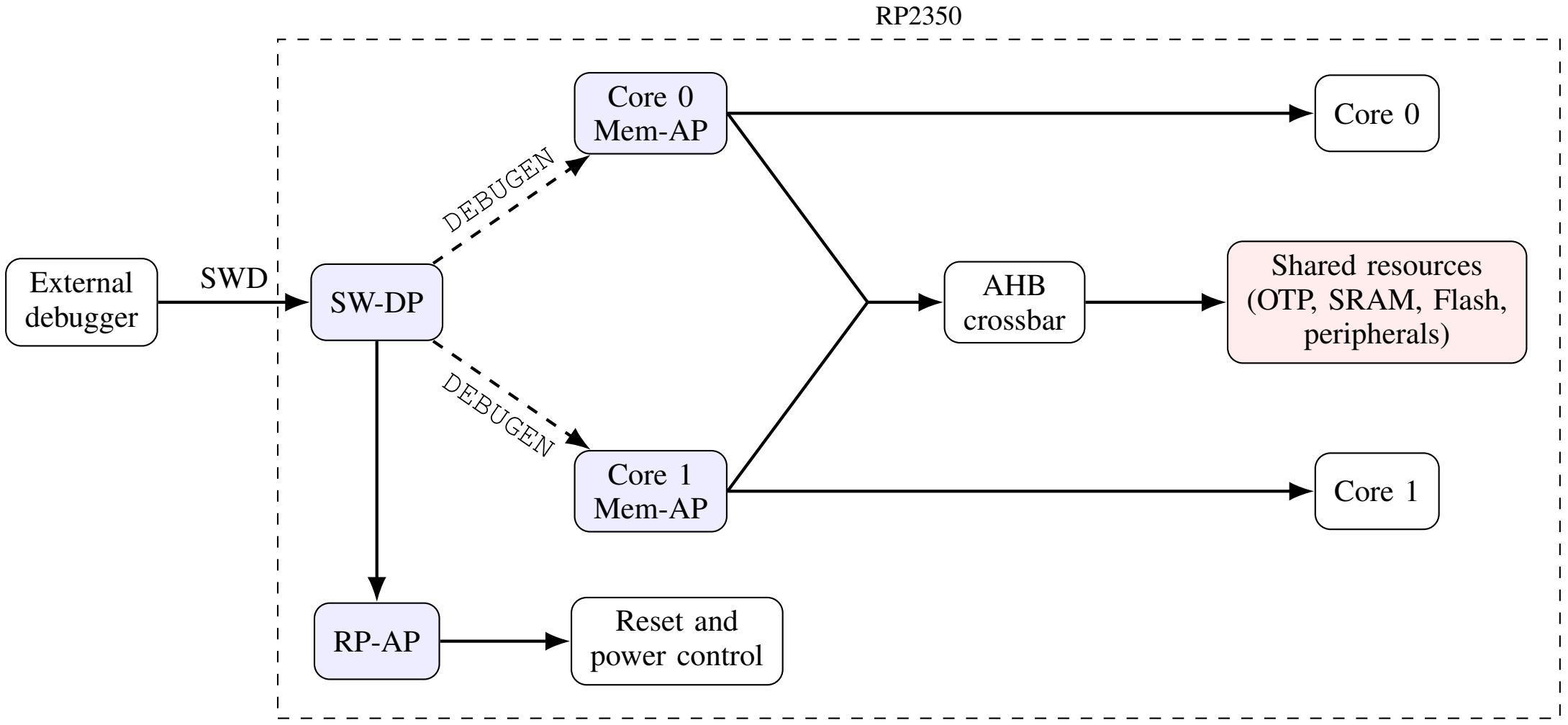
\begin{figure*}[!tb]
\centering
\begin{tikzpicture}[
  font=\footnotesize,
  >=Latex,
  box/.style={draw, rounded corners, align=center, minimum height=7mm,
              inner xsep=5pt},
  port/.style={box, fill=blue!7},
  securebox/.style={box, fill=red!7},
  arrow/.style={->, thick},
  gated/.style={->, thick, dashed}
]
  \node[box] (dbg) {External\\debugger};
  \node[port, right=14mm of dbg] (swdp) {SW-DP};
  \node[port, above right=10mm and 12mm of swdp] (mem0) {Core~0\\Mem-AP};
  \node[port, below right=10mm and 12mm of swdp] (mem1) {Core~1\\Mem-AP};
  \node[port, below=24mm of swdp] (rpap) {RP-AP};
  \node[box, right=22mm of swdp, xshift=24mm] (crossbar) {AHB\\crossbar};
  \coordinate[left=7mm of crossbar] (ahbjoin);
  \node[securebox, right=13mm of crossbar] (shared)
        {Shared resources\\(OTP, SRAM, Flash,\\peripherals)};
  \node[box] (core0) at (shared |- mem0) {Core~0};
  \node[box] (core1) at (shared |- mem1) {Core~1};
  \node[box, right=12mm of rpap] (reset) {Reset and\\power control};

  \draw[arrow] (dbg) -- node[above]{SWD} (swdp);
  \draw[gated] (swdp) -- node[above, sloped]{\reg{DEBUGEN}} (mem0);
  \draw[gated] (swdp) -- node[below, sloped]{\reg{DEBUGEN}} (mem1);
  \draw[arrow] (swdp) -- (rpap);
  \draw[thick] (mem0.east) -- (ahbjoin);
  \draw[thick] (mem1.east) -- (ahbjoin);
  \draw[arrow] (ahbjoin) -- (crossbar.west);
  \draw[arrow] (mem0.east) -- (core0.west);
  \draw[arrow] (mem1.east) -- (core1.west);
  \draw[arrow] (crossbar) -- (shared);
  \draw[arrow] (rpap) -- (reset);

  \node[draw, dashed,
        fit=(swdp)(mem0)(mem1)(rpap)(crossbar)(core0)(core1)(shared)(reset),
        inner sep=3mm, label={[font=\scriptsize]above:RP2350}] {};
\end{tikzpicture}
\caption{RP2350 debug architecture. Dashed links mark \reg{DEBUGEN} overrides
of the persistent \reg{CRIT1.DEBUG\_DISABLE} policy. Enabled Mem-APs access
shared resources and control their cores; the RP-AP remains available for
rescue reset. The RISC-V debug port is omitted because secure boot selects Arm.}

\label{fig:debug-architecture}
\end{figure*}

Each RP2350 processor socket supports an Arm Cortex-M33 or a Hazard3 RISC-V
core. Secure boot selects Cortex-M33 and verifies image signatures against
key fingerprints stored in OTP~\cite[Secs.~3.9 and~10.1.1]{rp2350-datasheet}.
A dedicated Redundancy
Coprocessor hardens critical boot code using encoded checks, sequence
counters, stack canaries, and panic-stop behavior~\cite{rp2350-security}.

Debug access follows the Arm CoreSight architecture, in which a debug port
selects access ports (APs) that reach on-chip resources~\cite{arm-adiv6}. The
RP2350 implements a Serial Wire Debug Port (SW-DP). Each core has a Mem-AP
that, when enabled, can access memory and peripherals and control its
processor through core-debug registers~\cite{arm-cortex-m33-trm}\cite[Sec.~3.5.2]{rp2350-datasheet}. The
Raspberry~Pi-specific RP-AP remains accessible when processor debug is
disabled. It exposes reset, power, rescue, and debug-key controls but cannot
issue bus transactions~\cite[Secs.~3.5.1 and~3.5.10]{rp2350-datasheet}
(Figure~\ref{fig:debug-architecture}).

The core~0 and core~1 Mem-APs, at \reg{0x2000} and \reg{0x4000}, issue
AHB5 transactions; the RP-AP is at \reg{0x80000}~\cite[Secs.~3.5.2 and~3.5.10.1]{rp2350-datasheet}.
Armv8-M TrustZone separates Secure and Non-secure access. Secure code and data are protected from Non-secure code.
Every bus transaction carries a security attribute that identifies the world that issued it, allowing peripherals to reject Non-secure access.
A debugger falls under the same rule. \reg{HNONSEC} marks an access as Non-secure.

\subsection{OTP Policy and Runtime Controls}

OTP is one-time-programmable memory built from antifuse bits. It holds the device's lifecycle configuration, boot keys and access policies.
Two of its mechanisms matter here.

\emph{Critical flags} are boolean settings that are read by hardware before any software, such as secure boot and debug disable.
Each flag is stored redundantly across 8 consecutive OTP rows, and hardware treats the flag as set if at least three of the eight copies are set~\cite[Sec.~13.4]{rp2350-datasheet}.
\reg{CRIT1.DEBUG\_DISABLE} disables all debug access ports except the RP-AP~\cite[Sec.~3.5.9.1]{rp2350-datasheet}. It can be overridden at runtime by the Secure-world mutable \reg{DEBUGEN} register.
Its bits are described in Table~\ref{tab:debugen}~\cite[Sec.~13.9, Table~1353]{rp2350-datasheet}. Restoring core~1 Secure debug therefore
requires bits~2 and~3, corresponding to the mask \reg{0xc}.

\begin{table}[t]
\caption{Functional \reg{DEBUGEN} bits. Bits~31:9 and 7:4 are
reserved~\cite[Sec.~13.9, Table~1353]{rp2350-datasheet}.}
\label{tab:debugen}
\centering
\begin{tabularx}{\columnwidth}{@{}c l >{\raggedright\arraybackslash}X@{}}
\toprule
Bit & Field & Effect \\
\midrule
0 & \bitname{PROC0} & Enable the core~0 Mem-AP \\
1 & \bitname{PROC0\_SECURE} & Authorize core~0 Secure transactions \\
2 & \bitname{PROC1} & Enable the core~1 Mem-AP \\
3 & \bitname{PROC1\_SECURE} & Authorize core~1 Secure transactions \\
8 & \bitname{MISC} & Enable the CoreSight CTI and the RISC-V Debug
    Module APB-AP \\
\bottomrule
\end{tabularx}
\end{table}

Each \reg{DEBUGEN\_LOCK} bit prevents software writes to its corresponding
\reg{DEBUGEN} bit until reset~\cite[Sec.~13.9, Table~1354]{rp2350-datasheet}.
It is documented as a protection against accidental software
writes~\cite[Sec.~3.5.9]{rp2350-datasheet}.

\emph{Page locks} control access to the data. OTP is divided into 64 pages of 128 bytes, each with a lock setting that controls Secure and Non-secure access.
A lock has three states: read/write, read-only and inaccessible~\cite[Sec.~13.5.1]{rp2350-datasheet}. Unlike critical flags, lock bits are stored three times and resolved by majority vote.
Access to pages can be further restricted through runtime page locks, but not restored~\cite[Secs.~13.5 and~13.5.1]{rp2350-datasheet}.
The runtime lock of page~\(n\) is the \reg{SW\_LOCKn} register~\cite[Sec.~13.9, Table~1333]{rp2350-datasheet}.
\reg{PAGEn\_LOCK1} has three fields~\cite[Sec.~13.10, Table~1420]{rp2350-datasheet}: \reg{LOCK\_S} manages access from Secure world, \reg{LOCK\_NS} from Non-secure and \reg{LOCK\_BL} can be used by bootloaders.

\subsection{RP2350 Hacking Challenge}

The RP2350 Hacking Challenge required recovery of a 128-bit OTP secret
protected by secure boot and access locks. Published attacks targeted
power-on sequencing, signature verification, reboot-mode selection, and OTP
reads using voltage, electromagnetic, and laser faults or direct antifuse
imaging~\cite{muench2025rp2350}. The only laser attack targeted the A2 boot ROM during signature verification~\cite{muench2025rp2350,courdesses2025laser}.
The A4 stepping strengthened the boot ROM against faults, while OTP macro changes mitigated a separate
power-disruption issue~\cite{rp2350-a4-announcement}.

The secret resides on OTP page~48. Persistent policy denies Non-secure access
(\reg{LOCK\_NS = INACCESSIBLE}) but permits Secure reads and writes
(\reg{LOCK\_S = READ\_WRITE}), encoded as
\reg{PAGE48\_LOCK1 = 0x3c3c3c}~\cite{rp2350-challenge}. At startup, signed
firmware sets \reg{SW\_LOCK48 = 0b1111}, denying both Secure and Non-secure
access. Restoring debug after a normal startup therefore cannot expose
the secret.

\section{Experimental Setup and Threat Model}
\label{sec:setup}

We evaluated one RP2350A4 in two phases. During offline characterization,
our firmware toggled \reg{DEBUGEN} bits for PEM and reported register values during
LFI mapping. We then provisioned the device with the challenge's page-48
policy (Section~\ref{sec:background}), secure boot under our test signing key,
a 128-bit test secret, permanent debug disable, and glitch-detector
sensitivity setting~3 (maximum)~\cite{rp2350-challenge}. Signed challenge
firmware denied all page-48 access at startup.

In the online threat model, the attacker possesses the device and has access
to its physical SWD port, the exposed backside, and the calibrated laser setup.
The attacker knows neither the secret nor the signing private key. At the start
of the attack, secure boot and persistent OTP permissions are enforced, and
processor debug is disabled. Unsigned code cannot execute, and the attacker
cannot access memory or core registers through the core Mem-APs.

Reproducing the experiment requires destructive preparation for backside
access. The setup we used for the attack cost approximately USD~250,000.

\section{Differential PEM Localization}
\label{sec:pem}

We used differential PEM to locate emission associated with toggling
\reg{DEBUGEN} bits. We recorded a long-exposure image while the chip ran a loop
that repeatedly toggled a chosen set of bits.
We ran this procedure for different sets of bits to compare the emissions by subtraction.
We call the image obtained by subtraction and its related set of bits a contrast.
We studied five contrasts. Three compare the controls of each core, while two compare the two controls of the same core.
The difference images gave candidate regions for the subsequent laser scanning.

The device was connected to the Scaffold control
platform~\cite{scaffold}.
Table~\ref{tab:pem-config} summarizes the acquisition
parameters. The detector's thermoelectric cooler (TEC) was set to
\SI{-15}{\celsius}, the manufacturer's recommended setpoint for optimum
performance~\cite{raptor-night-ninox640-manual}. To keep the sensor near this
setpoint without blurring the images, we ran the fan on the TEC hot side
during a break between cycles and switched it off during each
acquisition~\cite{perrin2025betrayed}.

\begin{table}[t]
\caption{PEM acquisition parameters}
\label{tab:pem-config}
\centering
\begin{tabular}{@{}p{0.46\columnwidth}p{0.45\columnwidth}@{}}
\toprule
Parameter & Value \\
\midrule
Detector & Night Ninox 640 InGaAs, \(640\times512\) pixels \\
Objective & \(20\times\), \(\mathrm{NA}=0.5\) \\
Pitch in sample plane & \SI{0.742}{\micro\meter}/pixel \\
Duration per acquisition & \SI{100}{\second} \\
Cycles per contrast & 200 \\
\bottomrule
\end{tabular}
\end{table}

\subsection{Controlled Excitation and Acquisition}

An SRAM-resident loop toggled selected \reg{DEBUGEN} bits twice per iteration,
with interrupts and host communication disabled. The loop used the register's
atomic XOR alias at
\(\reg{DEBUGEN}+\reg{0x1000}\)~\cite[Secs.~2.1.3 and~13.1]{rp2350-datasheet}. Writing a
mask \(M\) to this address updates the register as
\(R\leftarrow R\oplus M\): each set bit in the mask toggles the
corresponding register bit. Two writes of the same mask therefore toggle the
selected bits and then restore their initial values.

This allowed the loop to toggle the register without reading it or
alternating the values carried by successive writes. Both stores used the
same address and data. Conditions \(A\) and \(B\) ran the same instruction
stream at the same address and rate, differing only in the mask \(M\).

The \reg{0x3}-\reg{0xc} contrast is the difference between the two core~0
controls and the two core~1 controls. The \reg{0x1}-\reg{0x4} contrast is
the difference between the general debug enables of the two cores, and
the \reg{0x2}-\reg{0x8} contrast is the same difference for the
Secure-access controls. The \reg{0x1}-\reg{0x2} contrast is the
difference between the two controls of core~0, and the
\reg{0x4}-\reg{0x8} contrast is the same difference for core~1.

\begin{figure*}[!tb]
  \centering
  \includegraphics[width=\textwidth]{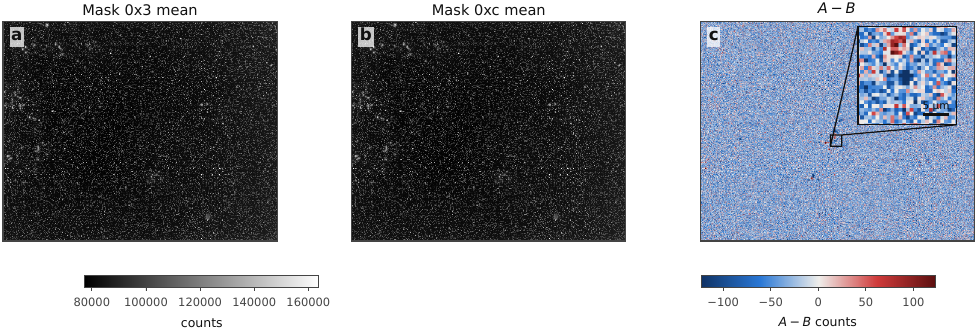}
  \caption{Mean images for masks \reg{0x3} and \reg{0xc} over all 200~cycles,
  and their difference. Subtraction suppresses optical and sensor structure
  shared by the two conditions.}
  \label{fig:pem-raw-vs-diff}
\end{figure*}

\begin{figure}[t]
\centering
\resizebox{0.98\columnwidth}{!}{%
\begin{tikzpicture}[
  font=\footnotesize,
  >=Latex,
  acquisition/.style={
    draw,
    minimum width=15mm,
    minimum height=9mm,
    font=\bfseries
  },
  conditionA/.style={acquisition, fill=red!7},
  conditionB/.style={acquisition, fill=blue!7}
]
  \node[conditionA] (a0) {A};
  \node[conditionB, right=1.5mm of a0] (b0) {B};
  \node[conditionB, right=9mm of b0] (b1) {B};
  \node[conditionA, right=1.5mm of b1] (a1) {A};

  \draw[->] ([xshift=-8mm,yshift=14mm]a0.center)
         -- ([xshift=12mm,yshift=14mm]a1.center) node[right,font=\scriptsize] {time};
  \coordinate (break) at ($(b0.east)!0.5!(b1.west)$);
  \draw[densely dashed] ([yshift=5mm]break) -- ([yshift=-5mm]break);
  \node[font=\scriptsize,above=6mm of break] {fan on};

  \node[fit=(a0)(b0), inner sep=0pt] (cycle0) {};
  \node[fit=(b1)(a1), inner sep=0pt] (cycle1) {};
  \node[fit=(a0)(a1), inner sep=0pt] (pair) {};

  \node[font=\scriptsize, below=8mm of cycle0,align=center]
       {cycle \(c\): slot effect \(-s_c(p)\)};
  \node[font=\scriptsize, below=8mm of cycle1,align=center]
       {cycle \(c+1\): slot effect \(+s_{c+1}(p)\)};

  \draw ([yshift=-13mm]pair.south west)
        -- ++(0,-1.5mm)
        -- ([yshift=-14.5mm]pair.south east)
        -- ([yshift=-13mm]pair.south east);
  \node[font=\scriptsize, below=16mm of pair]
        {order-balanced pair};
\end{tikzpicture}%
}
\caption{Order-balanced PEM acquisition. Each box is \SI{100}{\second}, with
the fan off. The fan runs during the break between cycles.}
\label{fig:pem-cycles}
\end{figure}
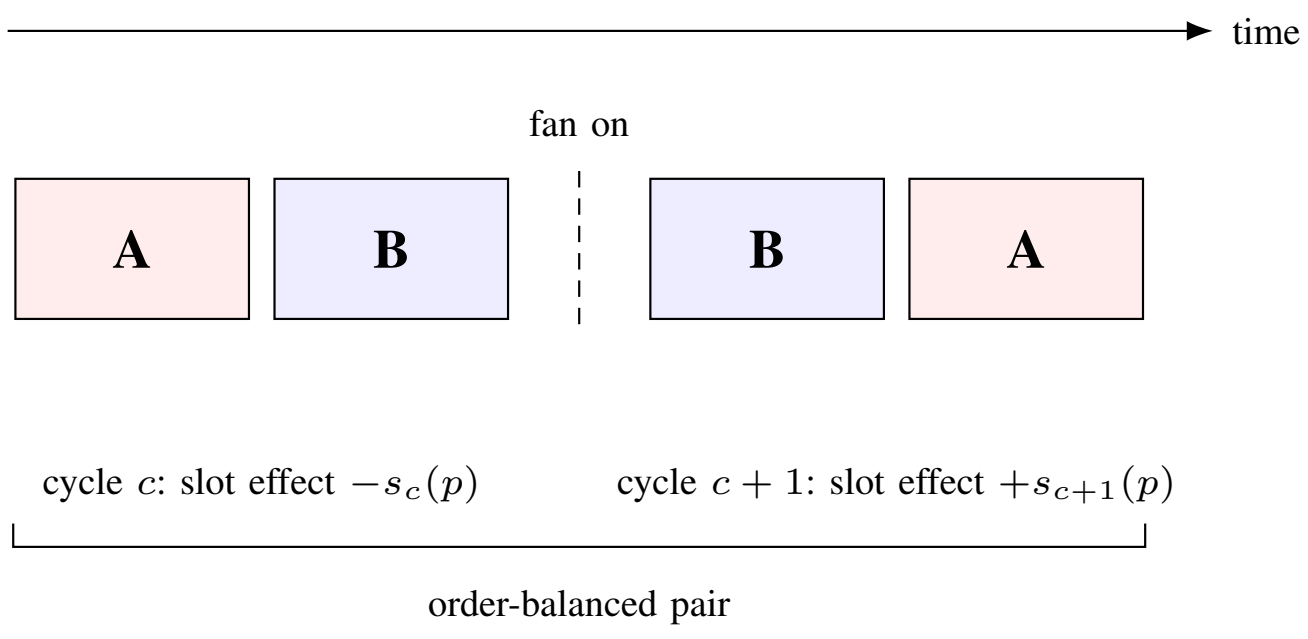

We acquired each contrast in a separate session of 200 cycles. Each cycle
contained one \SI{100}{\second} acquisition under each condition. We reversed
their order in successive cycles (Figure~\ref{fig:pem-cycles}), giving
an \(A,B,B,A\) or \(B,A,A,B\) sequence over two cycles, with a fan-on break
between the cycles. The total acquisition
time was approximately \SI{11.1}{\hour} per contrast.
Figure~\ref{fig:pem-raw-vs-diff} shows the mean images for the two conditions and their difference.

\subsection{Observed Image Variation}

The raw images shared a strong spatial pattern across masks
(Figure~\ref{fig:pem-raw-vs-diff}). Repeated acquisitions also showed changes
in overall brightness and spatially varying drift. To describe these
observations and the mask-dependent emission, we
write the intensity at pixel \(p\) under condition \(X\in\{A,B\}\) in cycle
\(c\) as
\begin{equation}
X_c(p)=E_X(p)+C(p)+o_{X,c}+d(p,\tau_{X,c})+\varepsilon_{X,c}(p),
\label{eq:pem-obs}
\end{equation}
where \(E_X(p)\) is emission that depends on the toggled mask. It can include
other mask-dependent circuit activity besides the register's storage elements.
The shared pattern \(C(p)\) includes emission unrelated to the mask and
fixed optical and sensor structure. The offset \(o_{X,c}\)
represents a frame-wide brightness shift, such as a change in the sensor
dark level caused by a change in its internal temperature. The term \(d(p,\tau_{X,c})\) captures spatially varying changes at
acquisition time \(\tau_{X,c}\). For example, a small shift in alignment or
focus can redistribute a bright feature across detector pixels, making some
pixels brighter and others dimmer. We did not isolate the
physical causes of these changes.
Finally, \(\varepsilon_{X,c}(p)\) collects measurement noise. This decomposition
motivates the processing below.

\begin{figure*}[!tb]
\centering
\includegraphics[width=0.86\textwidth]{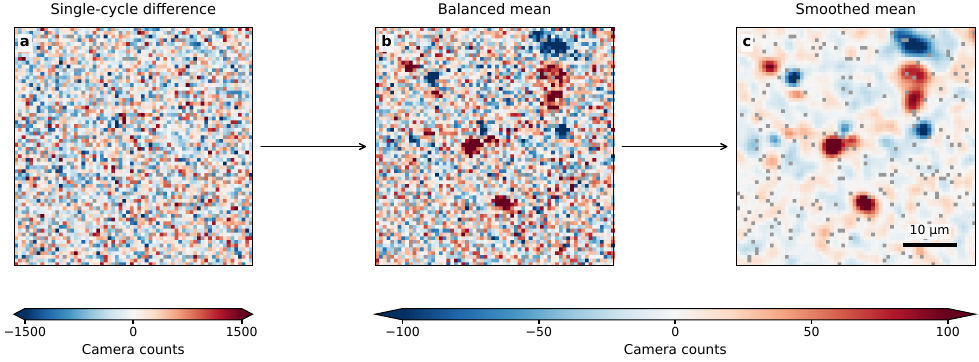}
\caption{Processing of one measured detector crop for \reg{0x3}-\reg{0xc} in
Region~I. (a) Centered difference from the earliest cycle retained in an
order-balanced pair. (b) Unsmoothed mean of all retained pairs.
(c) Mean after spatial averaging.}
\label{fig:pem-pipeline-toy}
\end{figure*}

\subsection{Processing and Map Construction}

\paragraph*{1) Remove Offsets and Shared Background}
We subtracted each acquisition's spatial median, then computed
\begin{equation}
\begin{aligned}
D_c(p)={}&A_c(p)-\operatorname{med}_{q}A_c(q)\\
            &{}-B_c(p)+\operatorname{med}_{q}B_c(q),
\end{aligned}
\label{eq:cycle-diff}
\end{equation}
where the median runs over all pixels \(q\). Centering removes \(o_{X,c}\);
differencing cancels \(C(p)\). The result contains the mask-dependent emission
difference, residual noise, and the difference between drift patterns at the
two acquisition times. Median centering also removes a spatially uniform constant from the emission difference.

\paragraph*{2) Filter Acquisition Cycles}
We screened for abrupt brightness changes between the two images in a cycle.
Let \(\bar A_c\) and \(\bar B_c\) be their raw whole-image means. We rejected a
cycle if \(\lvert\bar A_c-\bar B_c\rvert\) exceeded the across-cycle median
of this difference by ten scaled MADs (1.4826 times the median absolute
deviation~\cite{rousseeuw1993scale}). This flags a cycle in which one image
is unusually brighter or darker than its partner compared with typical
cycles.

\begin{figure*}[!b]
  \centering
  \includegraphics[width=\textwidth]{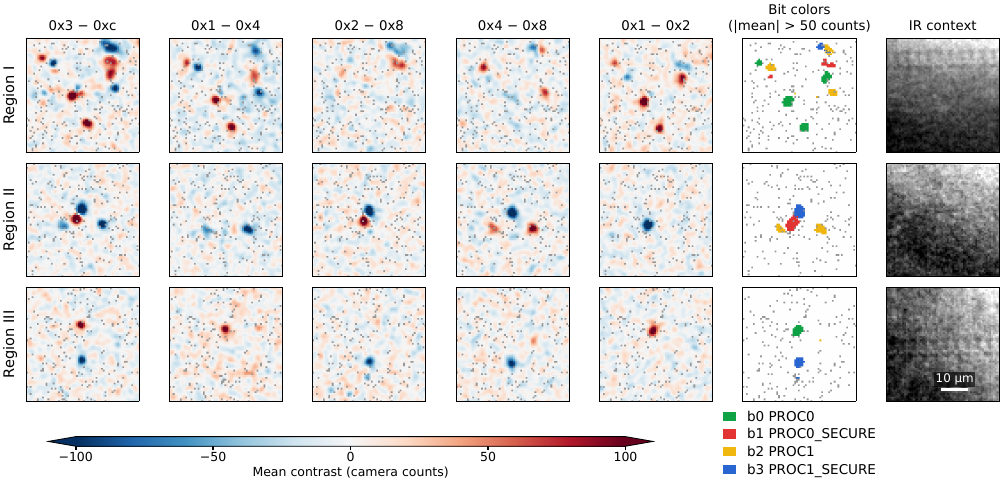}
  \caption{PEM contrasts in three regions. The first five columns show
  mean differences in camera counts on a common scale, each with its own
  validity mask. The sixth colors bits named by single-bit contrasts for easier interpretation.
  Gray indicates invalid pixels. The last column shows infrared context.}
  \label{fig:pem-bits}
\end{figure*}

\paragraph*{3) Pair Opposite Acquisition Orders to Reduce Slot Effects}
We observed a repeatable brightness change from the first to the second
acquisition of each cycle. Median centering removes its uniform component. Let \(s_c(p)\)
denote any remaining mask-independent, second-minus-first-slot change at
pixel \(p\) in cycle \(c\). For the \(A,B,B,A\) order, this effect enters
the two \(A-B\) differences as \(-s_c(p)\) and \(+s_{c+1}(p)\). We averaged
these differences and retained a pair only if both cycles passed screening.
The residual slot effect is
\begin{equation}
r_c(p)=\frac{s_{c+1}(p)-s_c(p)}{2}.
\label{eq:pem-slot-residual}
\end{equation}
It cancels when the slot effect repeats across cycles, even if the change
within a cycle is nonlinear. The \(B,A,A,B\) order reverses the residual's sign.

\paragraph*{4) Screen Pixels Within Each Session}
We excluded the one-pixel border and pixels that saturated or read zero in any retained pair. We then excluded pixels whose standard deviation across retained pairs lay outside the spatial median \(\pm\) ten scaled MADs.
Screening after pairing avoids rejecting stable pixels because of slot effects
that the pairing cancels.

\paragraph*{5) Spatial and Temporal Averaging}
We Gaussian-smoothed each retained pair with \(\sigma=1.13\)~pixels
(FWHM \SI{1.97}{\micro\meter}), normalizing over valid
neighbors~\cite{knutsson1993normalized}. We then averaged the retained pairs
per contrast (Figure~\ref{fig:pem-pipeline-toy}). The maps show localized
mean differences in three regions and served as visual guides for laser scanning
(Figure~\ref{fig:pem-bits}). The figure shows that spots from the \reg{0x3}-\reg{0xc} contrast reappear at
matching positions in the single-bit contrasts that share their bits, with the sign that each contrast predicts.
We treat this agreement as a consistency check, further indicating that these are not just emerging from noise.
Laser scans and register readback establish which
sites were fault-sensitive and which \reg{DEBUGEN} bits changed
(Section~\ref{sec:lfi}).

\section{Guided Laser Fault Injection}
\label{sec:lfi}

\begin{figure*}[!b]
  \centering
  \includegraphics[width=\textwidth]{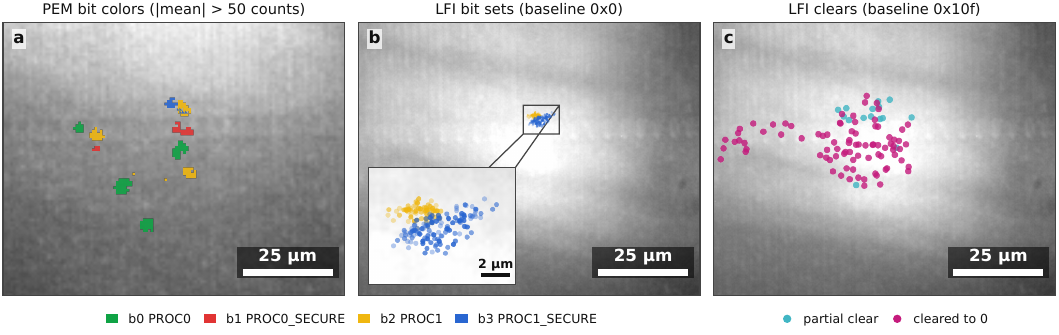}
  \caption{PEM and LFI maps of the same field of view. (a)~Bit-colored PEM
  summary from Figure~\ref{fig:pem-bits}, overlaid on the PEM infrared image.
  (b)~Positions where laser pulses produced sustained bit-setting changes in
  \reg{DEBUGEN} from
  an initial value of zero, overlaid on the LFI infrared image. The inset
  enlarges the two adjacent sites. (c)~Positions where pulses produced sustained
  bit-clearing changes from an initial value of \reg{0x10f}, on the same LFI image.
  These positions span \SI{53}{\micro\meter} by \SI{25}{\micro\meter}.}
  \label{fig:pem-lfi}
\end{figure*}

\subsection{Offline Characterization}

\subsubsection{Methods}

We scanned the region found with PEM with a \SI{980}{\nano\meter} laser and a
\(50\times\) objective. To locate bit-setting sites, we initialized
\reg{DEBUGEN} to zero and applied \SIrange{100}{120}{\nano\second} pulses
at 40--70\% of the laser's rated \SI{2.97}{\watt} maximum. A separate
bit-clearing scan started from \reg{0x10f}, using
\SIrange{10}{100}{\nano\second} pulses at 4--45\% of rated power.

Characterization firmware initialized the register before each pulse,
executed NOP instructions during injection, and reported \reg{DEBUGEN} and
\reg{DEBUGEN\_LOCK} over UART afterward, to prevent mislabeling transient faults. In the bit-setting scan, we accepted a nonzero
value as sustained only if a second read agreed after flushing the UART
buffers. We then restored zero by a software write, rebooting only if asserted
lock bits prevented the write.

To test whether software locks prevented laser-induced changes, we cleared
the enable bits and set the corresponding \reg{DEBUGEN\_LOCK} bits in
software.

\subsubsection{Results}

We measured approximately \SI{1.37}{\milli\meter\squared} of logic area on the exposed die. Guided by the three PEM candidate regions
(Figure~\ref{fig:pem-bits}), we focused the bit-setting scan on Region~I. The
scan starting from \reg{DEBUGEN = 0} covered
\SI{82.3}{\micro\meter\squared}, about 0.006\% of the measured logic area.
We identified two
sites approximately \SI{1.7}{\micro\meter} apart where pulses produced
sustained changes from zero to one in \bitname{PROC1} and \bitname{PROC1\_SECURE}
(Figure~\ref{fig:pem-lfi}b). No pulse set either \bitname{PROC0} bit.We did not determine why, and
did not extend the scan, because restoring \bitname{PROC1} and \bitname{PROC1\_SECURE} suffices to open core~1 Secure debug.

In the separate scan starting from \reg{DEBUGEN = 0x10f}, pulses cleared
bits at positions spanning \SI{53}{\micro\meter} by
\SI{25}{\micro\meter} (Figure~\ref{fig:pem-lfi}c), including both
\bitname{PROC0} bits. A pulse often cleared all implemented bits,
returning the register to zero.
Clearing sites were far more widespread
than setting sites.

Laser pulses still set enable bits when the corresponding software locks
were asserted. The lock bits remained set, and subsequent software writes
of zero did not clear the enables. Pulses could also set lock bits; we
observed no pulses that cleared them.

\subsection{Online Secure-Debug Restoration}

\subsubsection{Methods}

The online attack used \SI{100}{\nano\second} pulses at 40\% of rated power,
guided solely by SWD feedback. After each pulse, we reconnected and read
core~1's Mem-AP Control/Status Word (CSW), whose \reg{DeviceEn} and
\reg{SDeviceEn} fields indicate general and Secure access,
respectively~\cite{arm-adiv6}.

We targeted the calibrated \bitname{PROC1} site until \reg{DeviceEn} was
set, then switched to \bitname{PROC1\_SECURE}. If a pulse cleared the
general enable, we returned to \bitname{PROC1}. Once \reg{SDeviceEn} was
asserted, we confirmed Secure access by reading the Cortex-M33 CPUID through
a Secure transaction and checking it against the expected
value~\cite{arm-adiv6,arm-cortex-m33-trm}.

\subsubsection{Results}

Core~1 Secure access was confirmed by three observations: \reg{SDeviceEn}
was asserted, a Secure CPUID read succeeded, and \reg{DEBUGEN} read back as
\reg{0xc}. Access remained available without further pulses until a processor
cold reset. After calibration, the sequence restored access in as little as
\SI{5}{\second}.

Whenever \reg{DEBUGEN} remained at \reg{0xc},
\reg{DEBUGEN\_LOCK} read back as \reg{0x10f}, with all implemented lock bits
set. These faults enabled Secure debug and locked the enable bits, preventing
software from disabling debug until reset.

\section{End-to-End Compromise}
\label{sec:end-to-end}

As a negative control, we restored Secure debug after normal boot.
OTP reads remained denied with \reg{SW\_LOCK48 = 0b1111}. Debug restoration alone
therefore did not bypass the runtime lock. Recovering the secret required
restoring the page permissions that applied before user firmware tightened
the lock.

\subsection{Rescue-Reset Sequence}

The RP-AP's \reg{RESCUE\_RESTART} control resets the system and enters the
boot ROM's rescue path~\cite{rp2350-bootrom}. Rescue reset clears \reg{DEBUGEN} and reloads
runtime OTP locks from persistent policy~\cite[Secs.~3.5.8, 3.5.9
and~13.5]{rp2350-datasheet}. The boot ROM then leaves both processors waiting
without executing user firmware~\cite{rp2350-bootrom}. We used the following sequence to restore
Secure debug after reset and read page~48 before its runtime lock was tightened:
\begin{enumerate}
  \item We issued rescue reset through the RP-AP, reconnected over SWD, and
        confirmed that both core Mem-APs were disabled.
  \item We applied the SWD-guided sequence of Section~\ref{sec:lfi} until
        \reg{DEBUGEN} read back as \reg{0xc}.
  \item We enabled Secure transactions through core~1's Mem-AP and verified that page~48's runtime lock permitted Secure reads.
  \item We read the secret from the OTP.
\end{enumerate}
Because user firmware did not execute, we did not have to complete the
sequence within a limited window before page-lock initialization.

After rescue reset and debug restoration, OTP reads recovered the 128-bit test secret.

\Needspace{10\baselineskip}
\section{Discussion}
\label{sec:discussion}

Restoring Secure debug gives an attacker complete access to processor
state without bypassing secure-boot signature verification.
Firmware cannot fully mitigate this attack. It can deny the debugger
access to OTP through \reg{ACCESSCTRL}~\cite[Sec.~10.6]{rp2350-datasheet}
or monitor \reg{DEBUGEN} and trigger a cold reset when its value changes
unexpectedly, but both measures depend on firmware execution. An
attacker may halt the core before the monitor detects a change, and
rescue mode prevents the firmware from running and applying either
defense~\cite[Sec.~3.5.8]{rp2350-datasheet}.

The RP2350 protects its debug policy where it is stored, with redundant
encoding of critical flags and locks. At runtime, however, a single register,
\reg{DEBUGEN}, can override this protected policy.
Our attack targeted this weakest link in the enforcement chain.
Redundancy should therefore extend along the whole
chain, from OTP storage to the register that grants debug access.

\section{Conclusion}

Differential PEM guided laser scans that identified two sites where pulses
altered the bits controlling core~1 Secure debug on an RP2350A4. An SWD-guided
pulse sequence restored sustained access despite the asserted OTP debug-disable flag.
Combining this sequence with rescue reset recovered a 128-bit test secret
without modifying the stored OTP policy. This result shows that protecting
stored policy against faults is insufficient when an attacker can alter the
registers that enforce it.

\section*{Acknowledgment}
We disclosed the findings described in this paper to Raspberry~Pi
Ltd.\ on July~28, 2026. We thank their team for the prompt, professional
response and transparent approach to hardware security.

\bibliographystyle{IEEEtran}
\IEEEtriggeratref{23}
\bibliography{references}

\end{document}